\documentclass[11pt]{article}
\usepackage{blois,epsfig, amsmath}

\def\Journal#1#2#3#4{{#1} {\bf #2}, #3 (#4)}

\def\PLB{{\em Phys. Lett.}  B}
\def\PRL{\em Phys. Rev. Lett.}
\def\PRD{{\em Phys. Rev.} D}

\def\be{\begin{equation}}
\def\ee{\end{equation}}
\def\bea{\begin{eqnarray}}
\def\eea{\end{eqnarray}}

\begin{document}
\vspace*{4cm}
\title{SPIN LIGHT MODE OF MASSIVE NEUTRINO RADIATIVE DECAY IN MATTER}

\author{ A. GRIGORIEV \footnote{ax.grigoriev@mail.ru}}

\address{Skobeltsyn Institute of Nuclear Physics, Moscow State University,\\
119992 Moscow, Russia}

\author{ A. LOKHOV \footnote{lokhov.alex@gmail.com}}

\address{Department of Quantum Statistics and Quantum Field Theory, Moscow State University,\\
119992 Moscow, Russia}

\author{ A. STUDENIKIN \footnote{studenik@srd.sinp.msu.ru}}

\address{Department of Theoretical Physics, Moscow State University,\\
119992 Moscow, Russia}

\maketitle

\textbf{Introduction.} The calculations of the radiative neutrino
decay performed by several authors~\cite{Smi78} have shown that the
process characteristics are substantially changed if the presence of
a medium is taken into account. Note, the influence of the background
matter was considered only in the vertex (it is the influence on the
neutrino magnetic moment). Here we are going to discuss the impact of
the medium onto the state of neutrino itself in the process which is
similar to the so called spin light of neutrino
($SL\nu$)~\cite{LobStuPLB03}, a new mechanism of the electromagnetic
radiation the detailed quantum treatment of which was evaluated in a
series of our papers ~\cite{StuTerPLB05}. Lately, there appeared
several investigations~\cite{Stu06,GriShiStuTerTroGRAVCOS08} of the
same effect but in connection to the electron which can be the source
of the spin light of electron ($SLe$). The mechanism of $SL$ is based
on helicity states energy difference of the particle arising due to
weak interaction with the background matter, in addition the energy
deposit depends on the difference between the total energies of
neutrinos (or electrons) in the initial and final states. In neutron
matter, a more significant case is the $SL\nu$ process with
participation of antineutrinos and thus it is what we will study
here. However for the convenience in what follows we will still speak
about neutrinos.

We will consider the transition of one neutrino mass state $\nu_1$
into another mass state $\nu_2$ assuming that $m_1>m_2$, and restrict
ourselves with only these two neutrino species and accordingly with
two flavour neutrinos (like in the previous works). Since the
interactions of flavour neutrinos with neutron star matter, composed
predominantly of neutrons, are the same and governed by the neutron
density we will take equal interactions for the initial and final
massive neutrinos with the matter.

\textbf{Modified Dirac equation.} The system ``neutrino
$\Leftrightarrow$ dense matter" depicted above can be circumscribed
mathematically in different ways. Here we use the powerful method of
exact solutions developed in ~\cite{StuTerPLB05,Stu06}. In the frame
of this method neutrino states in matter are described by the
modified Dirac equation: $
\{i\gamma_{\mu}\partial^{\mu}-\frac{1}{2}\gamma_{\mu}(1+\gamma^{5})f^{\mu}-m\}\Psi(x)=0,
$ where for the case of unpolarized and stationary matter
$f^{\mu}=\frac{G_{F}}{\sqrt{2}}(n,\textbf{0})$. For the neutron star,
the matter density can reach the values of $n \approx 10^{37} \div
10^{40} cm^{-3}$. The corresponding energy spectrum of neutrino is
given by $ E_\varepsilon=\varepsilon\sqrt{(p-s\alpha
    m_{\nu})^{2}+m_{\nu}^{2}}+\alpha m_{\nu},$
where $\varepsilon = \pm 1$ defines the sign of the energy, $s$ is
the neutrino helicity, $p$ is the neutrino momentum and $\alpha =
\frac{1}{2\sqrt{2}}{G}_F\frac{n}{m_{\nu}}$. The exact form of the
solution of the modified Dirac equation
can be found in \cite{StuTerPLB05,Stu06}.

\textbf{Spin light mode of massive neutrino decay.} Using the
above-mentioned results we can now compute the process itself.
From the energy-momentum conservation law we get for the emitted
photon energy {\footnote {Recalling the results on the
$SL\nu$~\cite{LobStuPLB03,StuTerPLB05,Stu06} for the initial and
final neutrino states we take $s_1=-1, s_2=1$.}}
\begin{equation}\nonumber
    w=\frac{K(p+\Delta)-p\cos\theta+\sqrt{(K(p+\Delta
    )-p\cos\theta)^2-(K^2-1)(\Delta^2+2p\Delta
    )}}{K^2-1},
\end{equation} where $K=\frac{E-p \cos{\theta}+\alpha m_1}{\alpha
    m_1}$, $\Delta = \frac{m_1^2-m_2^2}{2\alpha m_1}$ and where
    $\theta$ is the angle between the direction of the initial
neutrino momentum $\bf p$ and the direction of the $SL\nu$ photon
emission. Using the exact neutrino wave functions in matter we
calculate the rate of the process,
   $ \Gamma_{\nu_{1} \nu_{2}}=\int_{0}^{\pi}\frac{d\Gamma}{d\cos\theta}d\cos\theta$.

\textbf{Results and discussion.} It is worth to investigate the
asymptotical behavior of the rate $\Gamma_{\nu_{1} \nu_{2}}$ in three
most significant limiting cases keeping only the first infinitesimal
order of small parameters. So we have,
\\
   $1) \ \ \Gamma_{\nu_{1} \nu_{2}}=4\mu^2\alpha ^3
m_1^3\Big(1+\frac{3}{2}
    \frac{m_1^2-m_2^2}{\alpha  m_1 p}+\frac{p}{\alpha  m_1}\Big),
    \ \ 1 \ll \frac{p}{m_1} \ll \frac{\alpha  m_1}{p}$ \
    (for ultrahigh density case),
\\
        $  2)\ \ \Gamma_{\nu_{1} \nu_{2}}=4\mu^2\alpha ^2 m_1^2 p
    \Big(1+\frac{\alpha  m_1}{p}+
    \frac{m_1^2-m_2^2}{\alpha
    m_1p}+\frac{3}{2}\frac{m_1^2-m_2^2}{p^2}\Big), \ \
    \frac{m_1^2}{p^2}\ll \frac{\alpha  m_1}{p} \ll 1$ \ (for high
    density case),
\\
    $  3)\ \ \Gamma_{\nu_{1} \nu_{2}}\sim\mu^2 \frac{m_1^6}{p^3},
    \ \
    \frac{\alpha  m_1}{p} \ll \frac{m_1}{p} \ll 1, \ m_1 \gg m_2$\
    \ (for quasi-vacuum case).
\\
Note that $\mu$ is the effective transition magnetic moment of the
neutrino~\cite{GiuStu09}.

The results 1) and 2) obtained within the generalization of the
method of exact solutions to the case when the masses of the
initial and final neutrinos are different, in the limit $m_1=m_2$
reproduce exactly the results of the $SL\nu$
calculation~\cite{LobStuPLB03,StuTerPLB05}. The asymptotic
estimation 3) gives the contribution of the spin light mode to the
neutrino radiative decay rate.

 \textbf{Acknowledgements.} The authors (A.L. and
A.S.) are thankful to Jean Tran Thanh Van, Jacques Dumarchez and
Ludwik Celnikier for the kind invitation to attend the 21st
Rencontres de Blois and to all of the organizers for their
hospitality in Blois. The authors are also thankful to Carlo
Giunti for stimulating discussions.

\end{document}